\documentclass[epjCONF,columns]{svjour} 
\usepackage{graphics}
\usepackage[varg]{txfonts} 
\usepackage[latin1]{inputenc}
\usepackage{lineno}
\usepackage{epsfig}
\usepackage{xspace}

\session-title {2011 Hadron Collider Physics symposium (HCP-2011), Paris, France, 
November 14-18 2011}
\newcommand{\sigmatt} {\ensuremath{\sigma_{t\bar{t}}}\xspace}
\newcommand{\ttbar} {\ensuremath{t\bar{t}}\xspace}
\newcommand{\ppbar} {\ensuremath{p\bar{p}}\xspace}

\newcommand{\ljets} {\ensuremath{\ell+\mathrm{jets}}\xspace}
\newcommand{\ifb}{\mbox{\,fb$^{-1}$}}
\begin{document}
\title{{\textbf {$t\bar{t}$}} and single top cross sections at the Tevatron}
\author{Elizaveta Shabalina\inst{1} \fnmsep\thanks 
{\email{Elizaveta.Shabalina@cern.ch}} for the CDF and D0 collaborations }
\institute{Georg-August-Universit\"at G\"ottingen, Friedrich-Hund-Platz 1, 
D-37077 G\"ottingen, Germany}

\abstract{
We present a summary of the latest measurements of the top pair and single top 
cross sections performed by the CDF and D0 collaborations at the Fermilab Tevatron 
collider. 
} 
\maketitle
\section{Introduction}
\label{intro}
The Fermilab Tevatron collider ended its run on September 30, 2011 after 
delivering more than 10 \ifb of \ppbar collision data per experiment at   
$\sqrt{s}=1.96$~TeV. A large sample of top quarks collected by the CDF and D0 
experiments allows to perform precision measurements of their production 
which is predicted to occur within the standard model (SM) either in pairs via strong 
interactions or as single top events via electroweak interactions.
Such measurements represent an important test of the 
theoretical calculations which predict the \ttbar and single 
top production cross sections with a precision of 6\% to 8\% ~\cite{SMtheorytt}
and 5\%~\cite{SMtheorystop}, respectively. Precise measurements  
of top pair cross section (\sigmatt) in different \ttbar\ final states 
and single top production via different production mechanisms 
are highly desirable as they are sensitive to the non-SM particles 
that may appear in top quark production or decays. 

\section{Top quark pair production cross sections}
\label{sec:1}

Within the SM, top quark decays to a $W$ boson and a $b$-quark with almost 
100\% probability. Thus, $t\bar{t}$ events can create final states with 
two leptons (dilepton channel) if both $W$ bosons from top quark decay leptonically 
into $e\nu_e$, $\mu\nu_{\mu}$ or $\tau\nu_{\tau}$, single lepton (\ljets channel) 
with one $W$ boson decaying leptonically and another one hadronically into 
$q\bar{q}^{\prime}$, or no leptons (all hadronic channel) if both $W$ bosons 
decay hadronically.      

\subsection{\ljets channel}
\label{sec:2}
The \ljets channel provides the most precise measurements of the \ttbar production 
cross section due to its relatively large branching and manageable 
background dominated by the production of $W$ bosons in association with heavy 
and light flavor jets ($W$+jets). Two approaches are used to discriminate \ttbar 
signal from the background. 
The first one exploits differences in kinematic 
properties of signal and $W$+jets background. It relies on selecting a number of 
discriminating variables, typically related to different features of the events, 
such as angular distributions of the objects or their energy, and building a 
discriminant from these variables using one of the multivariate techniques. 
The cross section is then extracted from a binned maximum likelihood fit to data.  
The same 
idea is used to extract single top cross section with the difference that for the 
latter one has to include many more discriminating variables due to a very poor 
signal over background (S/B) ratio.        

The second approach makes use of the presence of a $b$-quark in top quark decay while 
the major backgrounds are dominated by light jets production. Thus the requirement of 
at least one jet per event be identified as a $b$-jet improves significantly  
S/B and allows to extract \ttbar cross section from a counting experiment.      
 
D0 performed \ttbar cross section measurement with 5.3 \ifb of data by selecting events 
with $\ge 2$ jets and splitting them into subsamples according to the number of  jets 
and $b$-tags \cite{d0xsljets}. In the background dominated 
subsamples, i.e., two jet events, three jet events with $< 2$ $b$-tags and 
events with at least four jets and no $b$-tags, 
Random Forest (RF) 
discriminant is built to separate signal from the background. In the signal 
dominated ones, with at least four jets and three jets with two $b$-tags, 
the $b$-tag 
counting is used. Cross section is extracted from the simultaneous fit to data of the RF 
discriminant and the number of $b$-tags distribution across different jet multiplicities 
constraining many systematic uncertainties which are included as nuisance parameters 
in the fit. 
The measured cross section yields 
$\sigma_{t\overline{t}} = 7.78^{+0.77}_{-0.64} {\rm (stat+syst+lumi)} \, {\rm pb}$, 
achieving a relative precision below 10\%.  
The latter is limited by the systematic uncertainties with the 
largest contribution from the determination of integrated luminosity (6.1\%). 
 
The CDF collaboration significantly reduces the dependence on the luminosity measurement 
and its associated large systematic uncertainty by exploiting the correlation between 
the luminosity measurements for $Z$ boson and \ttbar production \cite{cdfxsljets}. CDF analysis computes   
the ratio of the \ttbar to $Z$ boson cross section, measured using the same triggers and dataset, 
and multiplies this ratio by the precisely known theoretical $Z$ cross section, thus 
replacing the luminosity uncertainty with the smaller theoretical and 
experimental uncertainties on $Z$ cross section. Using this approach in the dataset of 
4.6~\ifb CDF collaboration measures cross section using two methods: 
by constructing a neural network (NN) discriminant based on kinematic information and by 
counting $b$-tags.  
Statistical combination of these two measurements yields the most precise 
top quark cross section measurement to date of $\sigma_{t\overline{t}}=7.70 \pm 0.52 \,{\rm pb}$ 
with a relative uncertainty of 7\%.
All cross sections quoted above assume top quark mass $m_t=172.5$ GeV. 
Figure~\ref{fig:xs} (left) shows the RF discriminant distribution for events with three
jets and no $b$-tags and Fig.~\ref{fig:xs} (middle) shows the jet multiplicity 
distribution for events with at least two $b$-tags in D0 data compared to 
the background model prediction and \ttbar signal contribution calculated using the 
measured cross section.  
Figure~\ref{fig:xs} (right) shows the NN output used to extract the 
\ttbar cross section in the kinematic analysis by CDF, for simulated signal and 
background events, and data.  
\begin{figure*}[htb]
\begin{center}
\begin{minipage}{17.0 cm}
\includegraphics[width=5.5cm]{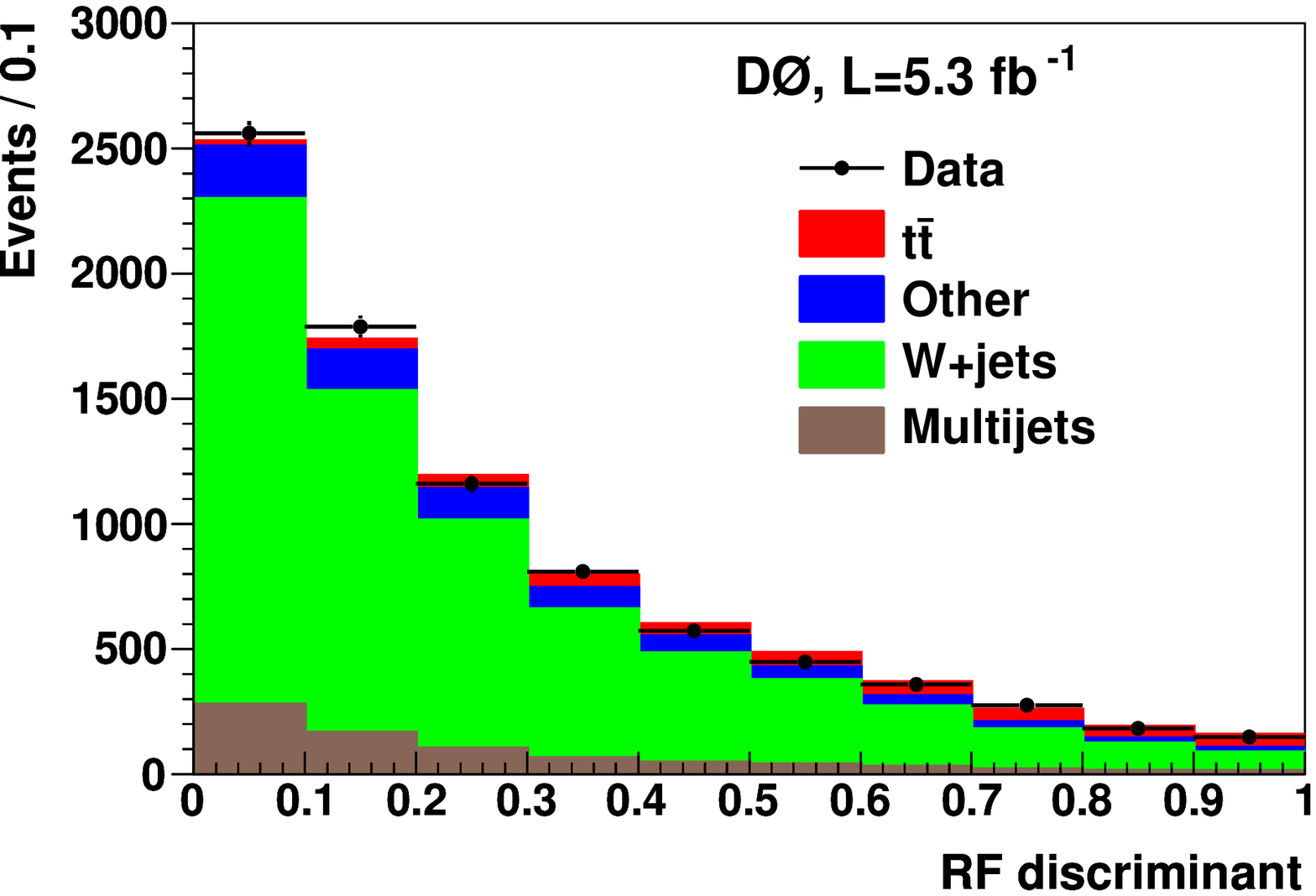}
\includegraphics[width=5.5cm]{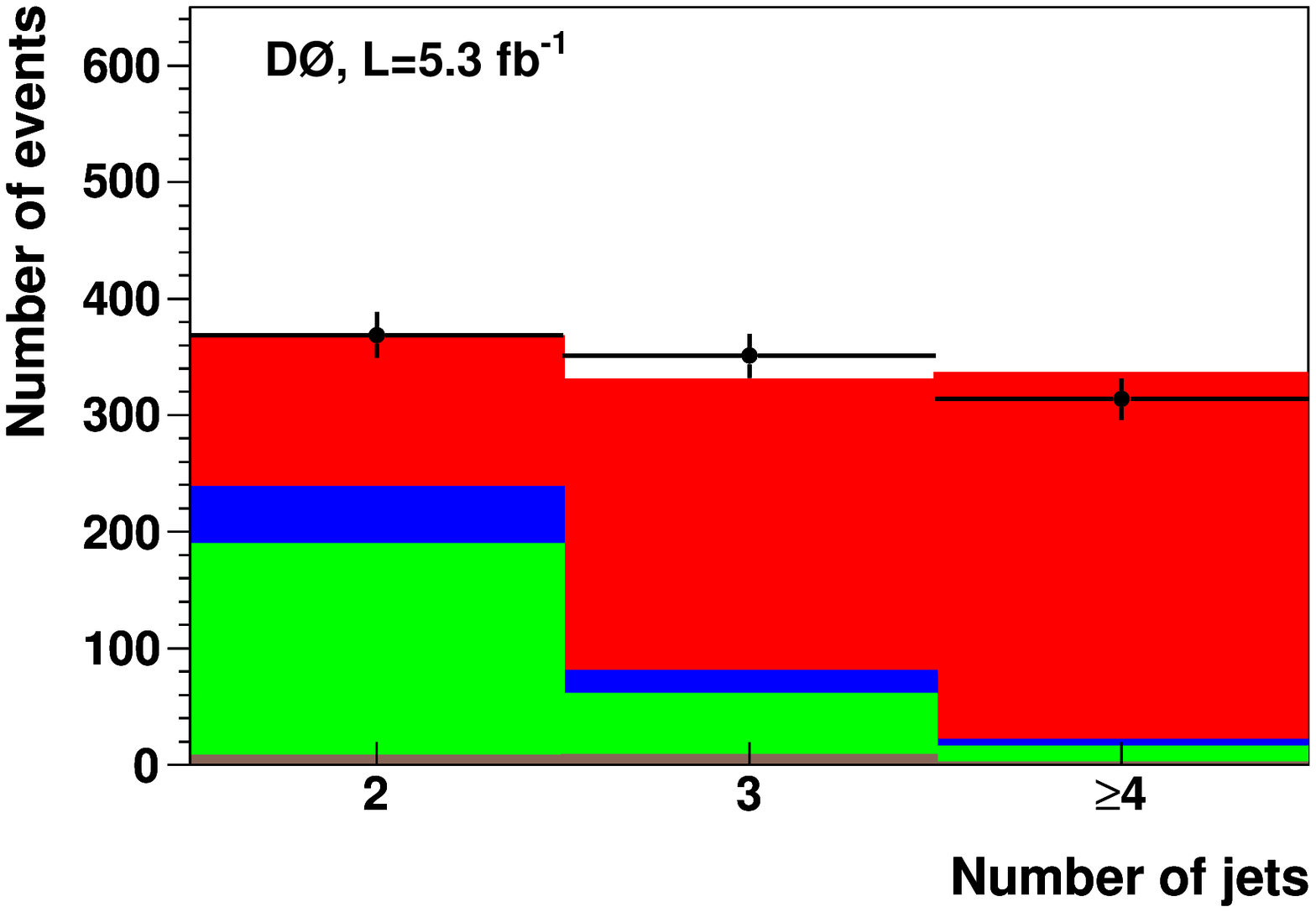}
\includegraphics[width=5.5cm]{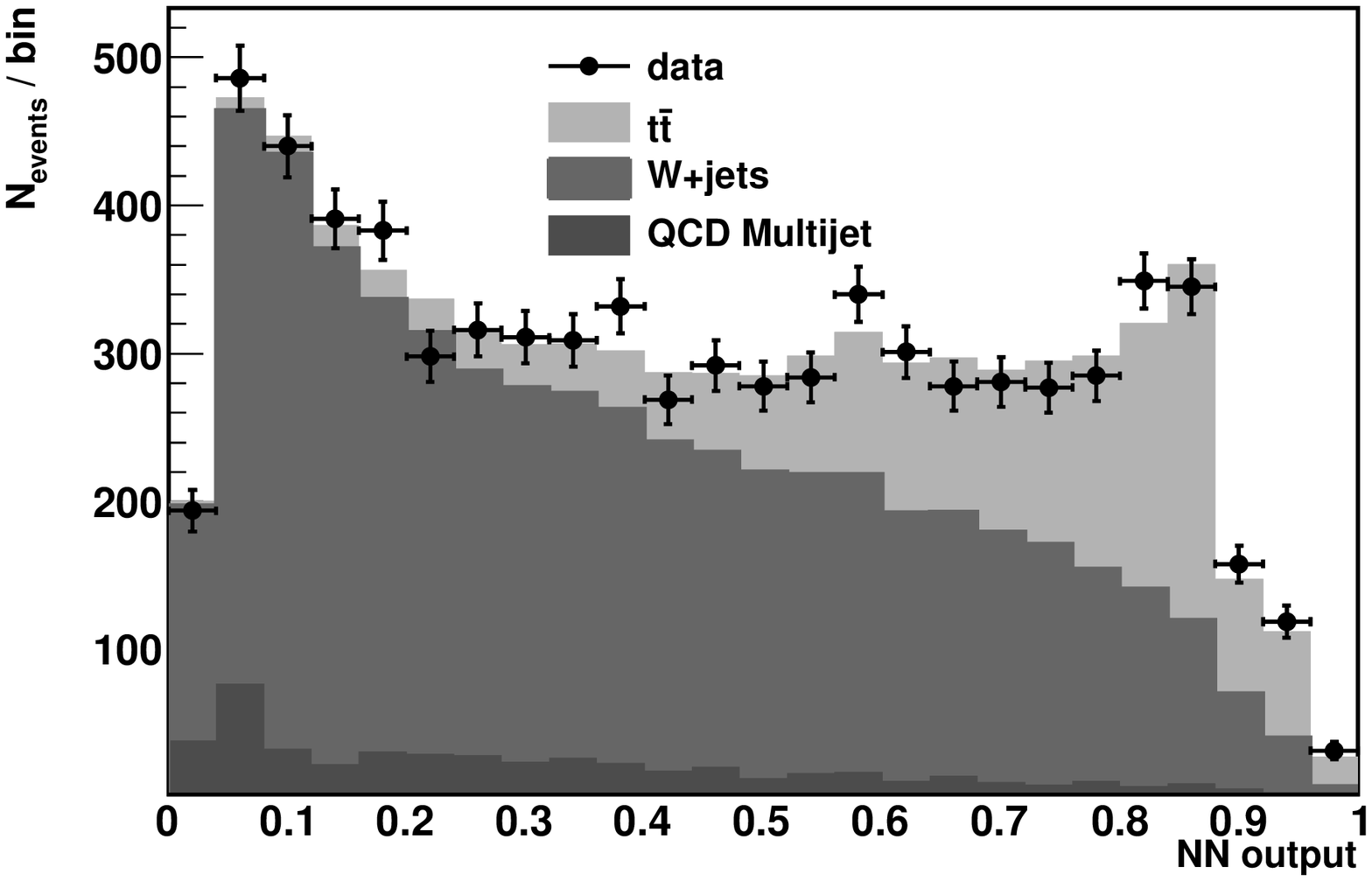}
\caption{RF discriminant output for events with three jets and no $b$-tags (left) and  
jet multiplicity spectrum of events with at least 2 $b$-tags (middle) from D0 analysis, NN output for events
with $\ge 3$ jets from CDF analysis (right). \label{fig:xs} }
\end{minipage}
\end{center}
\end{figure*}
\subsection{Dilepton channel}
\label{sec:2}
Low background, characteristic of the dilepton channel and dominated 
by the diboson and $Z$ boson production in association with jets, allows to  
use simple counting methods to extract \ttbar cross section. The disadvantage 
of this channel is its low branching fraction which made \sigmatt  
measurement statistically limited until recently.      
The CDF collaboration uses 5.1~\ifb of data to measure \sigmatt in the 
dilepton channel 
using 343 candidate events in the pretag sample and 137 events after 
$b$-jet identification requirement \cite{cdfxsdil}. Both measurements are in 
good agreement with each other and yield 
$\sigma_{t\overline{t}} = 7.40 \pm 0.58 {\rm (stat)} \pm 0.63 {\rm (syst)} 
\pm 0.45 {\rm (lumi)} \, {\rm pb}$ for the pretag analysis  
and $\sigma_{t\overline{t}} = 7.25 \pm 0.66 {\rm (stat)} \pm 0.47 {\rm (syst)} 
\pm 0.44 {\rm (lumi)} \, {\rm pb}$ for the $b$-tagged one. Statistical and systematic 
uncertainties are comparable for both approaches. The $b$-tagged analysis 
has a smaller systematic uncertainty due to a significantly lower backgrounds  
which have relatively large uncertainties and achieves a precision below 13\%.  
Figure~\ref{fig:cdfxsdil} shows jet multiplicity spectrum in the pretag sample. 
\begin{figure}[htb]
\begin{center}
\includegraphics[width=5.0cm]{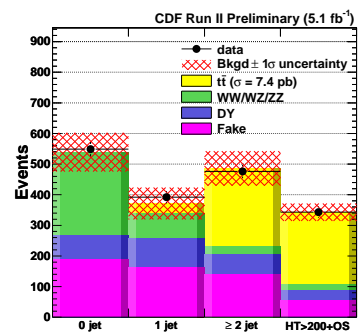}
\caption{Jet multiplicity spectrum of pretag events. \label{fig:cdfxsdil} }
\end{center}
\end{figure}

The D0 collaboration uses a more complicated new technique to measure \sigmatt in 
5.4~\ifb of data~\cite{d0xsdil}. In addition to events with at least two jets this analysis uses
one-jet events in $e\mu$ channel and extracts \sigmatt from the likelihood fit to 
the distriminant based on the D0 NN $b$-tagging algorithm. 
The algorithm combines information about the impact parameters of the tracks and 
variables that characterize the properties of the reconstructed secondary vertices 
in a single discriminant. The cross section is measured by simultaneously fitting 
the distributions of the smallest of the two $b$-tagging NN output values of the 
two leading jets in the four channels with systematic uncertainties included in 
the fit. This approach allows to constrain the uncertainties and achieve the best 
precision in the dilepton channel of 12\% resulting in   
$\sigma_{t\overline{t}} = 7.36^{+0.90}_{-0.79} {\rm (stat+syst+lumi)} \, {\rm pb}$.
The result is dominated by the systematic uncertainty with the largest uncertainty 
from luminosity of 0.57 pb. Figure~\ref{fig:d0xsdil} shows the output of the 
$b$-tagging discriminant for events with at least two jets in $e\mu$ channel in 
data and simulation.   
\begin{figure}[htb]
\begin{center}
\includegraphics[width=5.0cm]{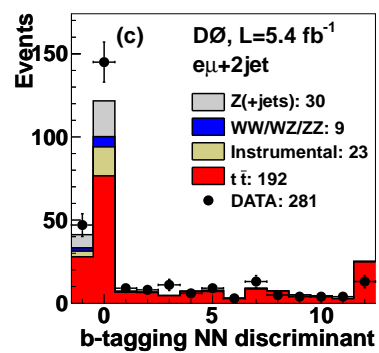}
\caption{Expected and observed distributions for the smallest $b$-tagging 
NN discriminant output of the two leading jets for $e\mu$ events 
with $\ge 2$ jets.  \label{fig:d0xsdil} }
\end{center}
\end{figure}
\vspace {-1.0cm}
\subsection{Combination of the cross sections}
Up to now \ttbar production cross sections were measured at the Tevatron 
in all channels except for the one with two hadronic taus in the final state using 
a variety of methods.   
All measurements are in agreement with the theoretical calculations. 
Figure~\ref{fig:d0xscombo} (\ref{fig:cdfxscombo}) shows a summary of the results 
from the D0 (CDF) experiment. A recent combination of dilepton and \ljets 
cross section measurements with 5.4\ifb by D0 yields  
$\sigma_{t\overline{t}} = 7.56^{+0.63}_{-0.56} \, {\rm pb}$ which corresponds 
to a relative uncertainty of (+8.3-7.4)\%~\cite{d0xsdil}. 
Combination of the best CDF measurements in 
different channels achieves relative uncertainty of 6.4\% and yields    
$\sigma_{t\bar{t}}=7.50\pm 0.48 \, {\rm pb}$~\cite{CDFxscombo} for   
a top mass of 172.5 GeV. 
\vspace {-1.0cm}
\begin{figure}[htb]
\begin{center}
\includegraphics[width=6.5cm]{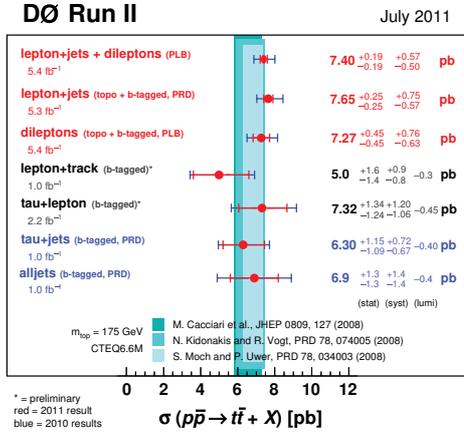}
\caption{Summary of the \sigmatt measurements by the D0 
experiment. \label{fig:d0xscombo} }
\end{center}
\end{figure}
\begin{figure}[htb]
\begin{center}
\includegraphics[width=7.0cm]{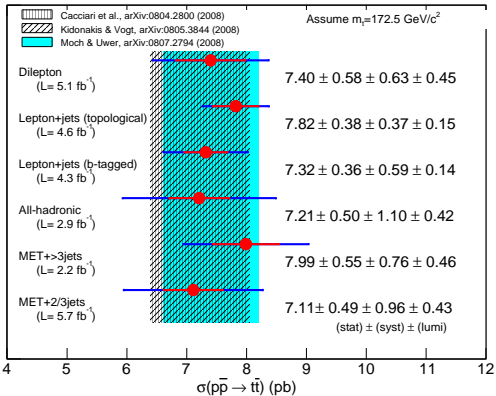}
\caption{Summary of the \sigmatt measurements by 
the CDF experiment. \label{fig:cdfxscombo} }
\end{center}
\end{figure}
\section{Production of single top quarks}
At the Tevatron
single top quarks are produced either by a $t$-channel exchange of a virtual $W$ 
boson which produces a top quark via interaction with a $b$-quark, or by an 
$s$-channel exchange of an off-shell $W$ boson which decays into a top and 
a $b$ quark. The third process, associated production of a $W$ boson and a top 
quark, has a negligible production rate at the Tevatron.    
Approximate NNLO calculation~\cite{SMtheorystop} 
predicts $t$-channel ($s$-channel) cross section to be $2.26\pm0.12$ pb ($1.04\pm0.04$ pb). 
Observation of the single top~\cite{st_obs} is difficult because of the 
relatively small cross section and very large background, and requires to use powerful 
multivariate techniques. 
\subsection{Cross section measurements}
Recent analysis of the D0 collaboration uses 5.4\ifb of data and measures combined $s+t$ channel 
cross section (assuming SM relative production rates for $t$- and $s$-channels)   
using three different multivariate discriminants constructed using 
Boosted Decision Trees~\cite{bdt}, 
Bayesian Neural Network (BNN)~\cite{bnn} and Neuroevolution of Augmented Topologies~\cite{neat}. 
The outputs of individual methods have correlation of 70\% and are combined 
using second BNN.  
The measured cross section is~\cite{d0stxs}  
$\sigma((s+t)) = 3.43^{+0.73}_{-0.74} \, {\rm pb}$ and has 21\% relative 
precision. Figure~\ref{fig:xsst} shows the output of the discriminant 
for a combined $s+t$ channel (left) and $s$-channel (middle) for the signal region. In these plots,
the bins are sorted as a function of the expected S/B ratio 
such that S/B increases monotonically within the range of the discriminant. 
Figure~\ref{fig:xsst} (right) shows the distribution of one of the most powerful 
discriminating variables,   
top quark mass, for the region of large value for signal discrimination (S/B~$>0.24$).  
Combined $s+t$ contribution is clearly visible in the plots while the signal 
presense in $s$-channel discriminant is not significant.     
\begin{figure*}[htb]
\begin{center}
\begin{minipage}{17.0 cm}
\includegraphics[width=5.5cm]{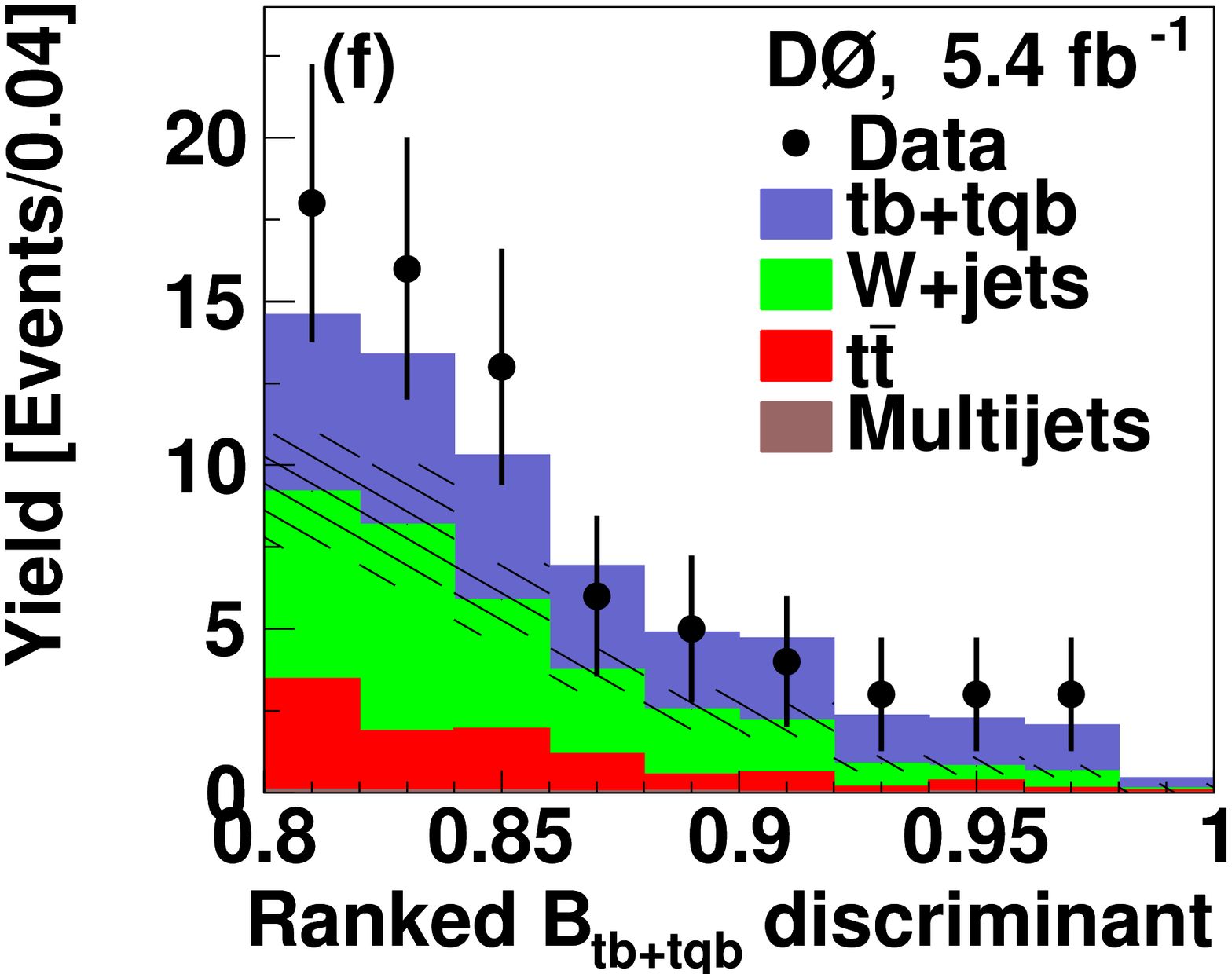}
\includegraphics[width=5.5cm]{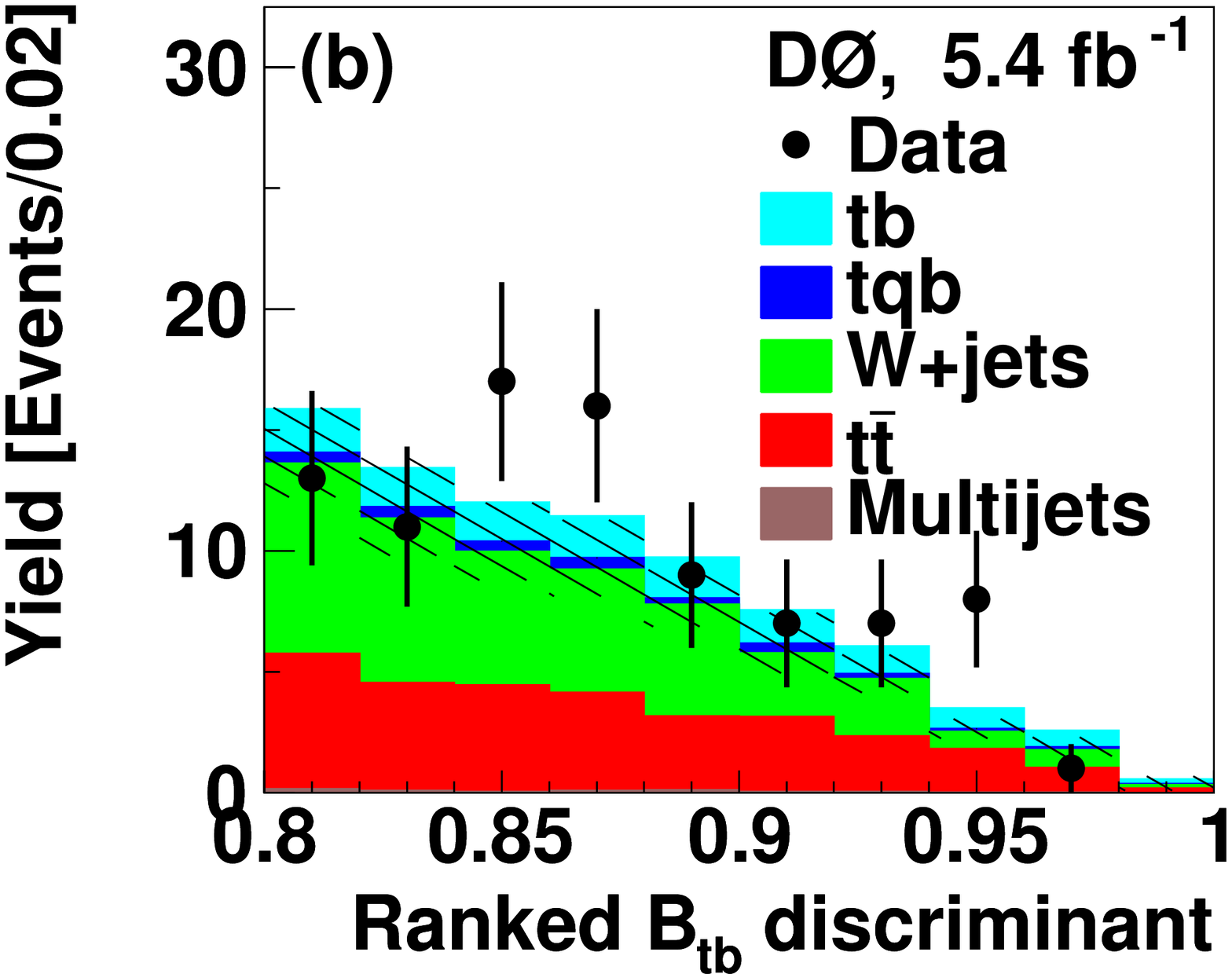}
\includegraphics[width=5.5cm]{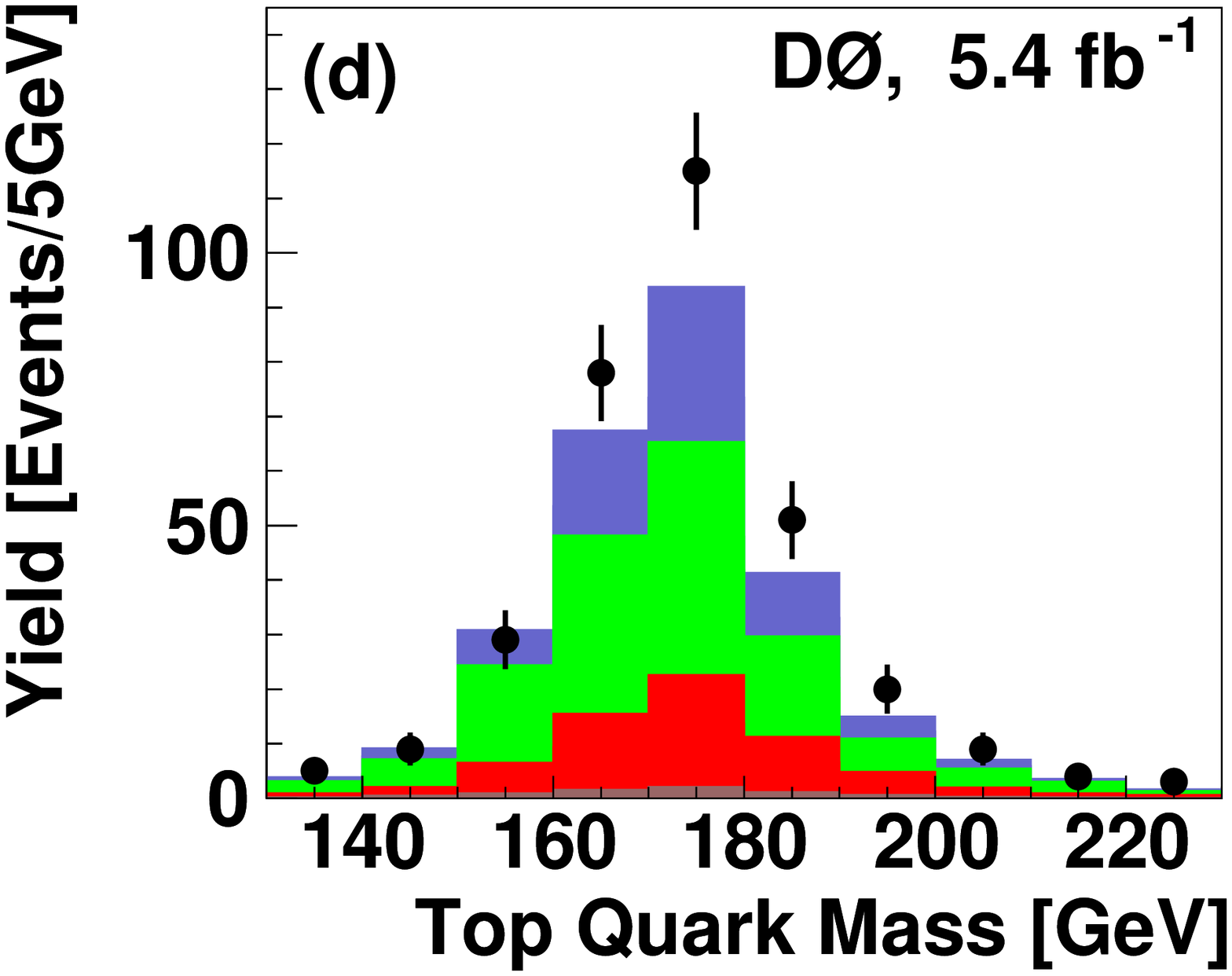}
\caption{Distributions of the $s+t$ discriminant (left) and $s$ discriminant (middle) 
in the signal region [0.8, 1] and distribution of the reconstructed top quark 
mass in a data sample with $S/B>0.24$ (right). \label{fig:xsst} }
\end{minipage}
\end{center}
\end{figure*}

Given that potential contributions from new physics scenarios can affect 
only one of the channels it is important to perform single top cross section 
measurements in $s$- and $t$-channels separately. Using same data set as for 
the combined $s+t$ measurement D0 collaboration performs model-independent 
measurement of $t$-channel single top production~\cite{d0stxst}. In this analysis, 
two-dimensional posterior probability density is constructed as a function 
of $t$- and $s$-channel cross sections without constraint on the relative rate of these 
contributions. Posterior distribution is integrated over $s$-channel axis to 
obtain one dimentional posterior probability density used to extract $t$-channel 
cross section. The latter is measured to be   
$\sigma(t) = 2.90 \pm 0.59 \, {\rm pb}$. The observed (expected) 
significance of the result is 5.5 (4.6) standard deviations.      
Cross section in $s$-channel is obtained in a similar way by integrating over 
$t$-channel axis: $\sigma(s) = 0.98 \pm 0.63 \, {\rm pb}$. 

The most precise measurement in the $s$-channel is obtained by the CDF collaboration 
$\sigma(s) = 1.8^{+0.7}_{-0.5} \, {\rm pb}$ in the single top 
observation analysis using 3.2\ifb of data~\cite{cdfstprd} with the significance of the signal 
exceeding 3 standard deviations.  
\section{Measurements of $|V_{tb}|$}
Single top production allows to probe $Wtb$ interaction since its production 
rate is proportional to the $|V_{tb}|^2$ mixing matrix element. The latter 
can be measured directly without any assumption on the number of quark families 
and the unitarity of the CKM matrix. D0 collaboration uses combined $s+t$ cross 
section measurement to extract $|V_{tb}|$ assuming that top quarks decay exclusively 
into $Wb$ and that $Wtb$ interaction is $CP$-conserving and of $V-A$ type. These 
assumptions allow an anomalous strength of the left-handed $Wtb$ coupling ($f_1^L$). 
Assuming the measured single top cross section~\cite{d0stxs} 
$|V_{tb}f_1^L| = 1.02^{+0.10}_{-0.11}$. Uncertainties include contributions from the 
theoretical uncertainties on the $s$- and $t$-channel production cross sections.        
If $f_1^L$ is assumed to be unity and $|V_{tb}|$ in [0,1] region, a 
limit $|V_{tb}|>0.79$ 
at 95\% C.L. is extracted. 

A measurement of \sigmatt can be used to study $|V_{tb}|$ indirectly by measuring the 
ratio of branching fractions 
\begin{eqnarray*}
\label{eq:Rdef}
R = \frac{{ \cal B}(t \rightarrow Wb)}{{ \cal B}(t \rightarrow Wq)} & = &
\frac{\mid V_{tb}\mid^2}{\mid V_{tb}\mid^2 + \mid V_{ts}\mid^2 + \mid
V_{td}\mid^2} \; 
\end{eqnarray*}
with $q$ being a $d$, $s$, or $b$ quark. 
In the SM $R=1$, constrained by the unitarity of the CKM matrix. $R<1$ can indicate a 
presence of new physics, for example, existence of additional quark families. 
D0 experiment measures $R$ simultaneously with the \ttbar cross section using dilepton 
and \ljets events in 5.4\ifb of data~\cite{rb}. In the \ljets channel, counting of $b$-tagged events 
in data and simulation is used to distunguish between different decay modes of \ttbar 
pair. In the dilepton channel, the shape of the output of the NN $b$-tagging algorithm            
serves the same purpose. Figure~\ref{fig:rblj} demonstrates a change of the predicted number 
of events with different $b$-tag multiplicity in \ljets channel while 
Fig.~\ref{fig:rbdil} shows a change of 
the shape of NN output in the dilepton channel for various values of $R$ compared to data.         
Simultaneous measurement of \sigmatt and $R$ yields $\sigmatt = 7.74^{+0.67}_{-0.57} {\rm pb}$ 
and $R = 0.90 \pm 0.04$, the latter being the most precise measurement of $R$ to date. 
Assuming unitarity of $3 \times 3$ CKM matrix $|V_{tb}|>0.88$ at 99.7\% C.L.  
\begin{figure}[htb]
\begin{center}
\includegraphics[width=5.0cm]{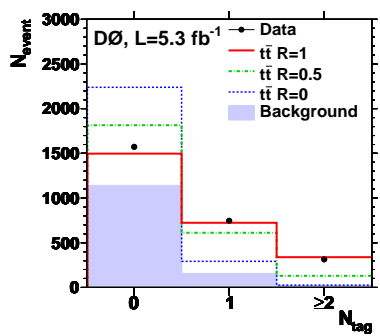}
\caption{Number of $b$-tagged jets in \ljets events with $\ge$ 4 jets. \label{fig:rblj} }
\end{center}
\end{figure}
\begin{figure}[htb]
\begin{center}
\includegraphics[width=5.0cm]{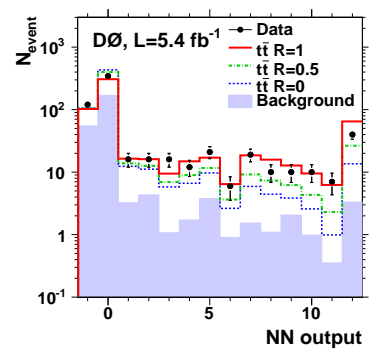}
\caption{Distribution of the minimum $b$-tag NN output of the jets of highest-$p_T$ 
in dilepton channel. \label{fig:rbdil} }
\end{center}
\end{figure}
\vspace {-1.0cm}
\section{Conclusions}
The Tevatron experiments provide precise measurements of the top pair 
production cross sections. Recent measurements of the electroweak single 
top production improve the uncertainty on the combined $s+t$ channel 
cross section and allowe to observe $t$-channel single top production.   
All results agree with the SM and challenge the precision of the 
theoretical calculations.

\end{document}
